\documentclass[prb,aps,twocolumn,floatfix]{revtex4}

\usepackage{graphicx,epsfig}
\usepackage{float}
\usepackage{dcolumn}
\usepackage{bm}
\usepackage{amssymb,amsmath,latexsym}

\begin{document}


\title{Spin-Wave Propagation in the Presence of Interfacial Dzyaloshinskii-Moriya Interaction}
\author{Jung-Hwan Moon$^1$, Soo-Man Seo$^1$, Kyung-Jin Lee$^{1,2 \dagger}$, Kyung-Whan Kim$^{3,4}$, Jisu Ryu$^3$, Hyun-Woo Lee$^3$, R. D. McMichael$^5$, and M. D. Stiles$^5$}
\affiliation{$^1$Department of Materials
Science and Engineering, Korea University, Seoul 136-701, Korea \\
$^2$KU-KIST Graduate School of Converging Science and Technology, Korea University, Seoul 136-713, Korea \\
$^3$PCTP and Department of Physics, Pohang University of Science and Technology, Pohang 790-784, Korea \\$^4$Basic Science and Research Institute, Pohang University of Science and Technology, Pohang 790-784, Korea \\$^5$Center for Nanoscale Science and Technology, National Institute of Standards and Technology,
Gaithersburg, Maryland 20899, USA}

\date{\today}

\begin{abstract}
In ferromagnetic thin films, broken inversion symmetry and
spin-orbit coupling give rise to interfacial Dzyaloshinskii-Moriya
interactions.  Analytic expressions for
spin-wave properties show that the interfacial
Dzyaloshinskii-Moriya interaction leads to non-reciprocal spin-wave
propagation, i.e. different properties for spin waves propagating in
opposite directions.  In favorable situations, it can increase the spin-wave attenuation
length. 
Comparing measured spin wave properties in
ferromagnet$|$normal metal bilayers and
other artificial layered structures with these calculations can provide a useful
characterization of the interfacial Dzyaloshinskii-Moriya interactions. 
\end{abstract}


\maketitle

\section{introduction}

Magnetic exchange is the root of magnetism.  Intraatomic exchange
stabilizes the magnetic moments and interatomic exchange tends to keep
the magnetization spatially uniform. Interatomic exchange is usually
symmetric in that the consequences of rotating the magnetization one way or the reverse are
equivalent. It looses that symmetry when the system is subject to both
spin-orbit coupling and broken inversion symmetry. The
antisymmetric component of the exchange interaction, known as Dzyaloshinskii-Moriya (DM)
interaction,\cite{Dzyaloshinskii, Moriya} can give chiral magnetic
orders such as spin spirals and skyrmions.\cite{Rossler2006, Uchida,
  Muhlbauer2009, Yi2010, Butenko2010, Yu2010Nat, Yu2010NatMat, Heinze,
  Huang} Understanding chiral magnetic order and its dynamics driven by magnetic fields or currents is currently of significant
interest in the field of spintronics.\cite{Jonietz2010, Tchoe, Yu2013NatComm,
  Thiaville, Iwasaki2013}

The DM interaction between two atomic spins $\mathbf{S}_i$ and $\mathbf{S}_j$ is
\begin{equation}\label{DMI}
{\cal H}_{\rm DMI} = -\mathbf{D}_{ij} \cdot (\mathbf{S}_i \times \mathbf{S}_j)
\end{equation}
where $\mathbf{D}_{ij}$ is the Dzyaloshinskii-Moriya vector, which is
perpendicular to both the asymmetry direction and the vector ${\bf
  r}_{ij}$ between the spins ${\bf S}_i$ and ${\bf S}_j$.  The DM
interaction can be classified into two classes depending on the type
of inversion symmetry breaking,\cite{Fert2013} i.e. bulk and
interfacial DM interactions corresponding to lack of inversion
symmetry in lattices and at the interface, respectively. The bulk DM
interaction has been studied mostly for B20 structures such as
MnSi,\cite{Muhlbauer2009} FeCoSi,\cite{Uchida, Yu2010Nat}
FeGe,\cite{Yu2010NatMat, Huang} etc. For the bulk DM interaction,
$\mathbf{D}_{ij}$ is determined by the detailed symmetry of the
lattice structure. On the other hand, the interfacial DM interaction,
which is the main focus of this work, occurs at all magnetic
interfaces.  It can be particularly strong at the interface between a
ferromagnet and a normal metal having strong spin-orbit coupling. The
DM interaction can be modelled by a 3-site exchange between two atomic
spins with a neighboring atom having a spin-orbit
coupling.\cite{Fert1980} It has been investigated for epitaxial
ferromagnet$|$heavy metal bilayers such as Mn/W,\cite{Bode2007,
  Ferriani2008} Fe/Ir,\cite{Heinze} and Fe/W.\cite{Heide2008,
  Udvardi2009, Zakeri2010, Zakeri2012}

Recently Chen \textit{et al.} reported that magnetic domain walls in
epitaxial Fe/Ni/Cu(001) structures are N\'{e}el walls and the domain
wall chirality is opposite to that of Ni/Fe/Cu(001)
structures.\cite{Chen2013} Such behavior is expected for an
interfacial DM interaction. The interfacial DM interaction in these
structures may be not as large as that of structures having a heavy
metal, but is still large enough to affect magnetic textures, which
can in turn modify magnetization dynamics
substantially.\cite{Thiaville} Furthermore, recent experiments on
current-driven domain wall motion suggest that the interfacial DM
interaction exists in sputtered Pt/CoFe/MgO\cite{Beach} and
Pt/Co/Ni\cite{Parkin} structures, and plays an important role in
domain wall motion. Since sputtered thin films consist of small grains
with different lattice orientation, the contributions from the bulk DM
interactions tend to cancel and only the interfacial DM interaction
contributions remain effective. In this respect, understanding the
interfacial DM interaction in sputtered thin films is important not
only for the fundamental understanding of topologically protected
nanomagnetic structures,\cite{Skomski} but also to the development of
spintronic devices based on domain walls.\cite{Ono,Ohno,Klaui}

\begin{figure}[ttbp]
\begin{center}
\psfig{file=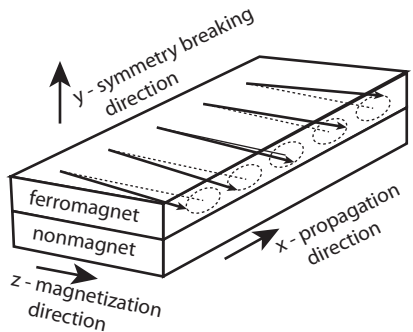} \caption{\label{fig:geom}
Geometry under consideration in this paper.  } 
\end{center}
\end{figure}

Translating the DM interaction in Eq.~(\ref{DMI}) to a continuum model
with magnetization direction $\hat{\bf m}$, and symmetry breaking in
the $\hat{\bf y}$ direction, the DM energy density is
given by
  \begin{eqnarray}
    \label{eq:intDM}
    {E}_{\rm DM} &=& -D\Bigg[ 
      (\hat{\bf x}\times\hat{\bf y}) \cdot\left(\hat{\bf m}\times
        \frac{\partial\hat{\bf m}}{\partial x}\right)
\nonumber\\ && ~~~~~~~
      +(\hat{\bf z}\times\hat{\bf y}) \cdot\left(\hat{\bf m}\times
        \frac{\partial\hat{\bf m}}{\partial z}\right)\Bigg]
  \end{eqnarray}
In this paper, we consider the case where the equilibrium
magnetization lies along the $\hat{\bf z}$ axis, in the plane of the
film.  We also restrict our attention to the case of spin waves
propagating in the $x$-direction so that ${\bf m}$ varies only in the
x-direction, and the second term in Eq.~(\ref{eq:intDM}) is zero (see
Fig.~\ref{fig:geom}).

A net DM interaction is present in any trilayer structure when the
first nonmagnetic layer supplies a spin-orbit coupling, the
middle layer is a ferromagnet, and the third layer is
nonmagnetic, but different from the first layer to break
symmetry.  Since the observation of efficient domain wall motion in
bilayers and trilayers is correlated with the conditions for a strong
DM interaction,\cite{Beach,Parkin} it is useful to study various
artificial structures to find large interfacial DM interactions. The
spin wave properties we present below provide a useful probe of the DM
interactions in these systems.

In this work, we compute analytical expressions for asymmetric
spin-wave propagation induced by the interfacial DM interaction.
There have been several related studies on specific systems. Udvardi
and Szunyogh predicted that the DM interaction gives rise to
asymmetric spin-wave dispersion depending on the sign of wavevector,
based on first principles calculations for
Fe/W(110).\cite{Udvardi2009} Costa \textit{et al.} predicted that the
spin-wave frequency, amplitude, and lifetime differ depending on the
sign of wavevector, based on a multiband Hubbard model for
Fe/W(110).\cite{Costa} Zakeri \textit{et al.} reported a series of
spin-wave experiments based on the spin-polarized electron-loss
spectroscopy for a single crystalline Fe/W(110),\cite{Zakeri2010,
  Zakeri2012} consistent with these predictions. Cort\'{e}s-Ortu\~{n}o
and Landeros developed a spin-wave theory for bulk DM
interaction,\cite{Landeros} where they demonstrated that the spin-wave
dispersion is asymmetric with respect to wavevector inversion.

We focus on the influence of the interfacial DM interaction on
spin-wave properties and pay attention to asymmetric spin-wave
attenuation and excitation amplitude with respect to wavevector inversion. We
provide analytic expressions for asymmetric dispersion, attenuation
length, and amplitude of interfacial DM spin-waves. In Sec. II, we
present spin-wave theory in the presence of the interfacial DM
interaction. Section III gives comparisons between analytic
expressions and micromagnetic simulations. We summarize our work in
Sec. IV.

\section{spin-wave theory}

We begin with a quantum spin-wave theory to find the contribution of
the interfacial DM interaction to the dispersion. Quantum spin-wave
theory for the symmetric exchange interaction is well
established,\cite{SWBook} and shows that the exchange interaction
results in $k^2$-dependence of the dispersion for small wavevector
$k$. Here we focus on the interfacial DM interaction in a
one-dimensional spin system. The interfacial DM interaction
Hamiltonian is given as
\begin{eqnarray}\label{Hamil_DM}
{\cal H}_{\rm DM} &=& -2 \frac{D_0}{\hbar^2} \sum_{j,\delta} \hat{\bf z} \cdot ( {\bf S}_{j} \times {\bf S}_{j+\delta}) \\ \nonumber
&=& \frac{D_0}{i \hbar^2} \sum_{j,\delta} (S_j^+ S_{j+\delta}^- - S_j^- S_{j+\delta}^+),
\end{eqnarray}
where $D_0$ is the DM energy, $S_j^+ (= S_{jx}+i S_{jy})$ and $S_j^-
(= S_{jx}-i S_{jy})$ are the spin raising and lowering operators. We
treat the case where the equilibrium magnetization direction is along
$\mathbf{D}_{ij}$ because this configuration exhibits the strongest
spin-wave asymmetry.\cite{Udvardi2009} Based on the 
Holstein-Primakoff transformation\cite{HP} and assuming that the total
number of flipped spins in the system is small compared to the total
number of spins, $S_j^+$ ($S_j^-$) can be approximated as $\hbar
\sqrt{2s} a_j$ ($\hbar \sqrt{2s} a_j^+$) where $s$ is the total spin
on the site, and $a_j$ ($a_j^+$) is magnon annihilation (creation)
operator. Substituting these approximations into Eq.~(\ref{Hamil_DM})
gives
\begin{equation}\label{Hamil_DM2}
{\cal H}_{\rm DM} = \frac{2 s D_0}{i} \sum_{j,\delta} (a_j a_{j+\delta}^+ - a_j^+ a_{j+\delta}).
\end{equation}
Introducing the operators $a_k^+$ and $a_k$, which are the Fourier
transforms of the $a_j$'s, and summing over $j$, Eq.~(\ref{Hamil_DM2})
becomes
\begin{eqnarray}\label{Hamil_DM3}
{\cal H}_{\rm DM} &=& \frac{2 s D_0}{i} \sum_{\delta, k} (e^{-ik \delta}a_k  a_k^+ - e^{ik \delta}a_k^+  a_k ).
\end{eqnarray}
The contribution to the magnon energy in Eq.~(\ref{Hamil_DM3}) is 
\begin{eqnarray}\label{Hamil_DM4}
{\cal H}_{\rm DM}^{\rm magnon} = -4 s D_0 \sum_{k} \sin(ka) a_k^+  a_k 
= \sum_{k} \hbar \omega_k^{\rm DM} \hat{n}_k,
\end{eqnarray}
where $\hat{n}_k=a_k^+ a_k$ is the number operator for magnons
with wavevector $k$, $a$ is the lattice constant, and the DM
interaction contribution to the dispersion is given by
\begin{equation}\label{freq_DM}
\hbar \omega_k^{\rm DM} = -4sD_0 \sin(ka).
\end{equation}
For small $k$, Eq.~(\ref{freq_DM}) reduces to 
\begin{equation}\label{freq_DM2}
\hbar \omega_k^{\rm DM} = -4sD_0 ka,
\end{equation}
a contribution to the dispersion that is linear in $k$. This
antisymmetric contribution to the energy leads to asymmetric spin-wave
propagation, i.e. dependent on the direction of $k$.

A similar contribution arises in a classical theory of spin-waves in
thin films with an interfacial DM interaction. We consider small
amplitude spin-waves propagating along the $x$-axis in the
perturbative limit, where the equilibrium magnetization is in the
$z$-direction perpendicular to both the film thickness direction and
the spin-wave propagation direction, 
\begin{equation}\label{mapprox}
{\hat{\bf m} =  p \hat{\bf z} + {\bf m}_0 \exp[i(k x -\omega t)] \exp[-x/\Lambda]},
\end{equation}
where ${\bf m}_0 =(m_x, m_y, 0)$, $|{\bf m}_0| \ll 1$, $p=\pm 1$, and
$\Lambda$ is the spin-wave attenuation length. The spin-wave dynamics
is described by the Landau-Lifshitz-Gilbert (LLG) equation, 
\begin{equation}\label{LLG}
{\partial \hat{\bf m} \over \partial t}= -\gamma \hat{\bf m} \times
\mu_0 {\bf H}_{\rm eff} + \alpha \hat{\bf m} \times {\partial \hat{\bf m} \over
\partial t},
\end{equation}
where $\gamma$ is the gyromagnetic ratio and $\alpha$ is the damping constant.  The effective field ${\bf H}_{\rm eff}$ is
given as   
\begin{eqnarray}\label{Heff}
{\bf H}_{\rm eff} & = &   p H \hat{\bf z}+J \nabla^2 \hat{\bf m} \\ \nonumber
&-& D^* \left(\hat{\bf z} \times \frac{\partial \hat{\bf m}}{\partial x} \right)+ {\bf H}_{\rm dipole}, 
\end{eqnarray}
where $H$ is the external field, $J$ is $2A/\mu_0 M_{\rm s}$, $D^*$ is
$2D/\mu_0 M_{\rm s}$, $A$ is the exchange stiffness constant, $M_{\rm s}$ is the
saturation magnetization, ${\bf H}_{\rm dipole}$ (= $-\frac{M_{\rm s}}{4}(1-e^{-2 |k| d})
 m_x \hat{\bf x}- M_{\rm s} (1-(1-e^{-2 |k| d})/4)m_y \hat{\bf y})$\cite{DE, Arias}
is the dipolar field, the local
demagnetization field along the thickness direction is equal to $M_{\rm s}$ , and $d$ is the
film thickness. We note that ${\bf H}_{\rm dipole}$ consists of local
contribution (independent of $d$ and $k$) and nonlocal contribution
(dependent on $d$ and $k$). Inserting Eqs.~(\ref{mapprox}) and (\ref{Heff}) into
Eq.~(\ref{LLG}), and neglecting small terms proportional to $1/(k
\Lambda)^2$, $\alpha^2$, and $\alpha /(k \Lambda)$, gives
\begin{widetext}
\begin{equation}\label{OmegaGeneral}
\frac{\omega}{\gamma \mu_0} = \sqrt{(H+M_{\rm s}/4+J k^2)(H+3 M_{\rm s}/4+J k^2)-\frac{e^{-4 |k| d}M_{\rm s}^2}{16}(1+2 e^{2 |k| d})}+ p {D^*} k,
\end{equation}
and
\begin{equation}\label{LambdaGeneral}
\Lambda_\pm = \frac{1}{\alpha \omega} \left( 2 \gamma \mu_0
  J|k_\pm|+ \frac{\gamma \mu_0 M_{\rm s}^2 d e^{-4 |k_\pm| d}(1+ e^{2 |k_\pm| d})/8 \pm p D^* (\omega \mp \gamma \mu_0 p {D^*} |k_\pm| )}{H+M_{\rm s}/2+Jk_\pm^2} \right), 
\end{equation}
\end{widetext}
where the upper (lower) sign corresponds to the case $k>0$
($k<0$). 
The dispersion (Eq.~(\ref{OmegaGeneral})) is the sum of the terms
  in the square root, which is the
  dispersion in the absence of the DM interaction, and a term linear
  in $k$.
Therefore, the interfacial DM interaction generates a term
linear in $k$ in the dispersion (Eq.~(\ref{OmegaGeneral}))
 as in the
quantum spin-wave theory (Eq.~(\ref{freq_DM2})). As a result, the
wavevectors are different for propagation in different directions at a
fixed frequency $\omega$. The spin-wave attenuation length also
depends on the sign of $k$ when $D \ne 0$ (Eq.~(\ref{LambdaGeneral})).

In the large $k$ limit (i.e., exchange-DM spin-waves), one may neglect the nonlocal magnetostatic contribution so that Eqs.~(\ref{OmegaGeneral}) and~(\ref{LambdaGeneral}) reduce to 
\begin{equation}\label{Omega1}
\frac{\omega}{\gamma \mu_0} = \sqrt{(H+J k^2)(H+M_{\rm s}+J k^2)}+ p {D^*} k,
\end{equation}
and
\begin{equation}\label{Lambda1}
\Lambda_\pm = \frac{1}{\alpha \omega} \left( 2 \gamma \mu_0
  J|k_\pm| \pm \frac{p D^* (\omega \mp \gamma \mu_0 p {D^*} |k_\pm|)}{H+M_{\rm s}/2+Jk_\pm^2} \right). 
\end{equation}
On the other hand, in the small $k$ limit (i.e., magnetostatic-DM spin-waves) that is more relevant to experimental conditions, one may neglect the exchange contribution and assume $ |k_\pm| d \ll 1$ so that Eqs.~(\ref{OmegaGeneral}) and~(\ref{LambdaGeneral}) reduce to
\begin{equation}\label{Omega2}
\frac{\omega}{\gamma \mu_0} = \sqrt{H(H+M_{\rm s})}+\frac{M_{\rm s}^2 |k|d}{4 \sqrt{H(H+M_{\rm s})}}+ p {D^*} k,
\end{equation}
and
\begin{equation}\label{Lambda2}
\Lambda_\pm = \frac{1}{\alpha \omega} \left( \frac{\gamma \mu_0 M_{\rm s}^2 d /4 \pm p D^* (\omega \mp \gamma \mu_0 p {D^*} |k_\pm|)}{H+M_{\rm s}/2} \right). 
\end{equation}
In this small $k$ limit, one finds from Eq.~(\ref{Omega2})
that not only the interfacial DM interaction but also the dipolar
coupling generates a term linear in $k$. However, there is an
important difference. The interfacial DM 
interaction contribution changes its sign with respect to the
inversion of the wavevector $k$ or the magnetization direction $p$,
whereas the dipolar contribution does not. Due to this feature, one
can distinguish the interfacial DM interaction contribution from the
dipolar contribution unambiguously.

Not only are the wave vectors, Eq.~(\ref{OmegaGeneral}), and the decay lengths,
Eq.~(\ref{LambdaGeneral}) asymmetric, the
amplitudes of the spin wave are different when symmetrically excited.
We approximate the ratio of spin-wave amplitudes $\kappa$
(=$m_y^+/m_y^-$) by neglecting contributions from nonlocal dipolar
coupling, where the plus (minus) sign corresponds to $k > 0$ ($k < 0$). From the susceptibility, one finds
\begin{equation}\label{my}
m_y = \sqrt{\frac{H_x}{H_y}} \int_0^\infty \frac{dk}{2\pi} \frac{h_k H_y}{2 \omega_0 \delta \omega + i \Gamma} \propto \frac{1}{v_g}, 
\end{equation}
where $H_x=H+J k^2$, $H_y=H+M_{\rm s}+J k^2$, $h_k$ is the Fourier component of the driving field, $\omega_0=\gamma \mu_0 \sqrt{H_x H_y}$, $\delta \omega$ describes the frequency difference from the resonance frequency, $\Gamma$ describes the damping term, and $v_g$ is the group velocity. Thus, the spin-wave
amplitude ratio $\kappa$ is
\begin{equation}\label{AmpRatio}
\kappa = \frac{-p D^*+J k_-\frac{2 H+M_{\rm s}+2Jk_-^2}{\sqrt{(H+Jk_-^2)(H+M_{\rm s}+Jk_-^2)}}}{+p D^*+J k_+\frac{2 H+M_{\rm s}+2Jk_+^2}{\sqrt{(H+Jk_+^2)(H+M_{\rm s}+Jk_+^2)}}}. 
\end{equation}
This equation shows that the interfacial DM interaction
makes the spin-wave amplitude asymmetric depending on the sign of $k$
or $p$.

These results only hold when the DM interaction is not strong enough
to change the ground state of the magnetic configuration. Setting
Eq.~(\ref{OmegaGeneral}) to be zero and neglecting nonlocal dipolar coupling,
the threshold $D_{th}^*$ is 
\begin{equation}\label{Dth}
{D_{th}^*} = \sqrt{[2 H +M_{\rm s}+2 \sqrt{H(H+M_{\rm s})}]J}.
\end{equation}
When $D^* > D_{th}^*$, the ground state is not a single domain but
rather a chiral magnetic texture and our results may not
apply. However, to study a particular interface, one can reduce the
thickness-averaged effective $D^*$ below $D_{th}^*$ by simply
increasing the thickness of ferromagnet because the DM interaction in
sputtered thin films is an \textit{interface} effect. By doing so, one
can study the DM interaction associated with a particular interface in
an appropriate layered structure.

\begin{figure}[ttbp]
\begin{center}
\psfig{file=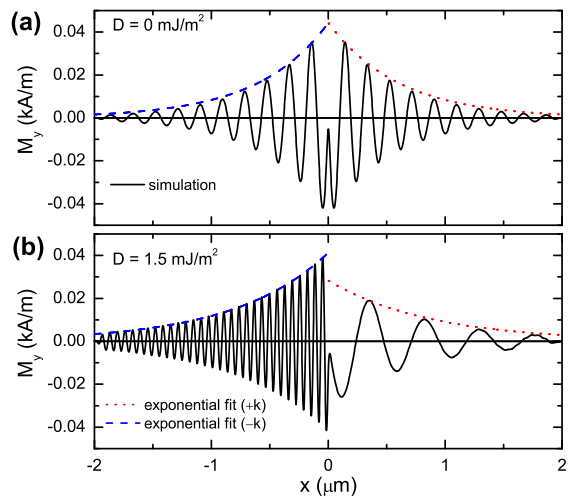,width=0.90\columnwidth} \caption{\label{fg1}
(color online) Snap-shot images of spin-wave distribution along the
$x$-axis for (a) $D$=0 mJ/m$^2$ and (b) $D$=1.5 mJ/m$^2$. The
spin-wave frequency $f$ is 11 GHz.  } 
\end{center}
\end{figure}

\section{numerical results and discussion}

To test the equations derived above, we perform
micromagnetic simulations in the large $k$ and small $k$ limits for a
semi-one dimensional system (i.e., the system is discretized along the
length direction, but assumed to be uniform along the
width and thin enough that there is no variation along the
thickness). In simulations for the large $k$ limit, we neglect the
nonlocal dipolar coupling but keep the local demagnetization field to
mimic a thin film geometry. On the other hand, we keep all micromagnetic
interactions in simulations for the small $k$ limit. We use the
damping constant $\alpha$ = 0.01, the saturation magnetization $M_{\rm
  s}$ = 800 kA/m, the exchange constant $A$ = 1.3 $\times$ 10$^{-11}$
J/m, the film thickness $d$ = 1 nm, and the film width of 20
$\mu$m. We use the external uniform field $\mu_0 H$ = 0.1 T (0.01 T)
and a unit cell size of 2 nm (5 nm) in simulations for the large
(small) $k$ limit. To excite spin-waves, we apply an ac field (0.1
mT)$\times \cos (2 \pi f t)$ to two unit cells at $x$ = 0. Therefore, 
the wavevector $k$ of spin-waves for $x > 0$ is positive whereas $k$
for $x < 0$ is negative. For a legitimate comparison between
theoretical and numerical results, we include absorbing boundary
conditions\cite{Berkov,Seo} at the system edges to suppress spin-wave
reflection.

Figure~\ref{fg1} shows snap-shot images of the spin-wave distribution
along the $x$-axis. Spin-waves are symmetric for $D$=0
(Fig.~\ref{fg1}(a)), whereas the wavelength, amplitude, and
attenuation length are all asymmetric depending on the propagation
direction for $D \ne 0$ (Fig.~\ref{fg1}(b)).

\begin{figure}[ttbp]
\begin{center}
\psfig{file=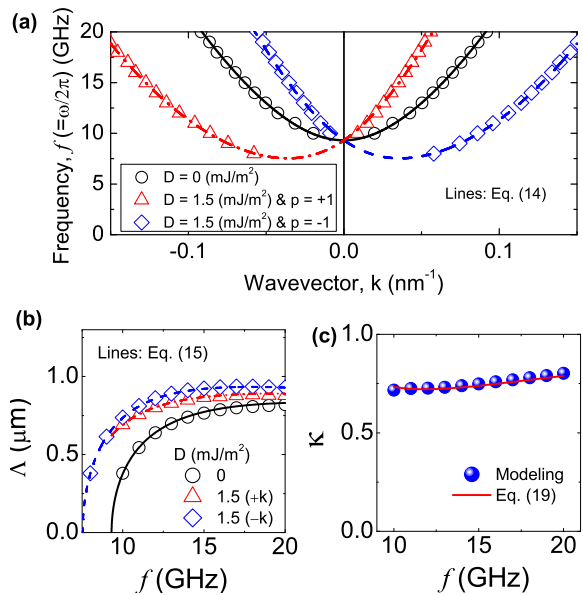,width=0.9\columnwidth} \caption{\label{fg2}
(color online) Asymmetric spin-wave propagation induced by an
interfacial DM interaction in the large $k$ limit: (a) Dispersion
relation. (b) Attenuation length $\Lambda$ as a function of the
frequency $f$. (c) Amplitude ratio $\kappa$ as a function of
$f$. Symbols and lines correspond to numerical and analytic results,
respectively. } 
\end{center}
\end{figure}

Figure~\ref{fg2} summarizes numerical results obtained in the large
$k$ limit. Numerical results of both spin-wave dispersion and
attenuation length are in agreement with analytic expressions
(Eqs.~(\ref{Omega1}) and~(\ref{Lambda1})). In this large $k$ limit, an
interesting observation is that $\Lambda$ for $D \ne 0$ is longer than
$\Lambda$ for $D = 0$ regardless of the propagation direction. After
some algebra with Eq.~(\ref{Lambda1}), one finds $\Lambda_\pm =
v_g^\pm F(k_\pm)/\alpha \omega$ where $v_g^\pm$ is the group velocity
and $F(k)=\sqrt{(H+M_{\rm s}+J k^2)(H+J k^2)}/(H+M_{\rm s}/2+J
k^2)$. Since $F(k)$ is a slowly varying function with the
wavevector, the attenuation legnth is mostly determined by the group
velocity. The interfacial DM interaction lowers the frequency gap
(i.e. the lowest allowed $\omega$) so that it increases the group
velocity at a given $\omega$ in the large $k$ limit, which in turn
increases the attenuation length compared to that with $D=0$. This
enhanced spin-wave attenuation length induced by the interfacial DM
interaction may be useful for applications based on spin-waves. We
note however that the damping constant $\alpha$ may also increase with
$D$, because the interfacial DM interaction is usually caused by
non-negligible spin-orbit coupling in the normal metal layer. In this
case, the damping may increase due to spin pumping
effects~\cite{Tserkovnyak2002} or interfacial Rashba spin-orbit coupling-related
spin-motive force~\cite{Kim2012, Tatara2013}.

Figure~\ref{fg2}(c) shows numerical results of the amplitude ratio
$\kappa$ as a function of the frequency $f$, in agreement with the analytic expression (Eq.~(\ref{AmpRatio})). We note that an asymmetry of
spin-wave amplitude has been observed when the spin-waves are excited
by a magnetic field generated by microwave antennas, and has been called
non-reciprocity of spin-waves.\cite{Bailleul2003, Amiri2007,
  Schneider2008, Demidov2009, Sekiguchi2010} In this case, the
amplitude asymmetry results from a non-reciprocal antenna-spin-wave
coupling, caused by the spatially non-uniform distribution of the antenna
field. However, we use reciprocal coupling in deriving
the analytic expressions and performing the numerical simulations, so
that the amplitude asymmetry shown in Fig.~\ref{fg2}(c) is purely due to
the interfacial DM interaction. This interfacial DM
interaction-induced amplitude asymmetry may find use in spin-wave
logic devices as proposed by Zakeri {\it et al.}.\cite{Zakeri2012} In
addition, it suggests a reexamination of the interpretation of
experiments reporting  non-reciprocal
antenna-spin-wave coupling.  These experiments have been done on relatively thin
structures, in which  interfacial DM
interaction may also be an important source of asymmetry.

\begin{figure}[ttbp]
\begin{center}
\psfig{file=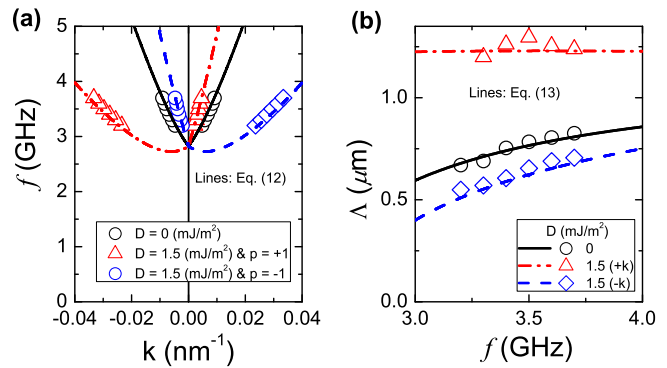,width=1.0\columnwidth} \caption{\label{fg3}
(color online) Asymmetric spin-wave propagation induced by an
interfacial DM interaction in the small $k$ limit: (a) Dispersion
relation. (b) Attenuation length $\Lambda$ as a function of the
frequency $f$. Symbols and lines correspond to numerical and analytic
results, respectively. } 
\end{center}
\end{figure}

Figure~\ref{fg3} summarizes numerical results obtained in the small
$k$ limit. Numerical results for both the spin-wave dispersion and
attenuation length agree with analytic expressions
(Eqs.~(\ref{Omega2}) and~(\ref{Lambda2})). In Fig.~\ref{fg3}(b), one
finds a difference in the attenuation length $\Lambda$ from
Fig.~\ref{fg2}(b). In the large $k$ limit, $\Lambda$ with $D \ne 0$ is
larger than $\Lambda$ with $D=0$ regardless of the sign of $k$. In
contrast, in the small $k$ limit with $pD > 0$, $\Lambda$ with $D \ne
0$ is larger than $\Lambda$ with $D=0$ for $k > 0$, whereas it is
smaller for $k<0$. This result is again related to the group
velocity. From Eq.~(\ref{Omega2}), one finds $v_g^\pm = v_g^0 \pm
\gamma \mu_0 p D^*$ where $v_g^0 = \gamma \mu_0 M_{\rm s}^2 d/4
\sqrt{H(H+M_{\rm s})}$ is the group velocity with $D=0$ in the small
$k$ limit. Therefore, for a sign of $k$, $\Lambda$ with $D \ne 0$ is
larger than $\Lambda$ with $D=0$ whereas for the other sign of $k$, it
is smaller.

Since the analytic expression of the dispersion
(Eq.~(\ref{OmegaGeneral})) is valid regardless of $k$, the strength of
interfacial DM interaction $D$ can be estimated experimentally by
measuring the frequency shift $\Delta f$ (=$|f_{+k,\pm p}-f_{-k,\pm
  p}|$=$|f_{\pm k, +p}-f_{\pm k,-p}|$), given as 
\begin{equation}\label{Freqshift}
\Delta f = \gamma \mu_0 D^* |k|/\pi .
\end{equation}
With the parameters $M_{\rm s}$ = 800 kA/m, $\gamma$ =
1.76$\times$10$^{11}$ T$^{-1}$s$^{-1}$, and $2 \pi/k$ = 1 $\mu$m,
$\Delta f$ is about 880 MHz for $D$ = 1 mJ/m$^2$ that is smaller than
the threshold value ($\approx$ 3.1 mJ/m$^2$ for the parameters used in
simulations, see Eq.~(\ref{Dth})). We note that propagating
spin-wave spectroscopy\cite{Vlaminck,BD1,BD2,Sekiguchi} can resolve
$\Delta f$ smaller than 20 MHz. 

Another interesting consequence of the
interfacial DM interaction is that the dispersion is asymmetric
depending not only on the wavevector direction but also on the
magnetization direction (i.e. the sign of $p$). This $k$- and
magnetization-direction-dependent asymmetry in the dispersion is
similar to the electron dispersion in a ferromagnet subject to Rashba
spin-orbit coupling.\cite{Park} We note that this is not an accident because the interfacial DM interaction is directly connected with the Rashba spin-orbit coupling at magnetic interfaces.~\cite{Kim2013}

\section{summary}

We theoretically study asymmetric spin-wave propagation induced by interfacial DM
interactions. We derive analytic expressions of
dispersion, attenuation length, and amplitude of interfacial DM
spin-waves and compare them with numerical results. The
frequency shifts induced by the interfacial DM interaction range from
MHz to GHz, which should be large enough to be resolved by 
state-of-the-art experimental tools such as propagating spin-wave
spectroscopy. Assuming that the damping does not change with the
interfacial DM interaction, the spin-wave attenuation length can
increase with increasing the interfacial DM interaction. The spin-wave
amplitude is asymmetric due to the interfacial DM interaction,
even without non-reciprocal coupling between antenna fields and
spin-waves. This asymmetric spin-wave propagation may be useful to investigate interfacial magnetic properties.

\section*{Acknowledgements}
This work was supported by the NRF (2010-0023798, 2011-028163, NRF-2013R1A2A2A01013188) and KU-KIST School Joint Research Program.

($\dagger$) Corresponding email: kj\_lee@korea.ac.kr



\end{document}